\documentclass[twocolumn,tighten]{aastex631}
\usepackage{amssymb}
\usepackage{amsmath}
\newcommand{\dxdy}[2]{{\frac{\partial{#1}}{\partial{#2}}}}
\newcommand{\dxdys}[2]{{\frac{\partial^{2}{#1}}{\partial{#2}^{2}}}}
\newcommand{\DxDy}

\newcommand{\bvf}{Brunt--V\"ais\"al\"a}
\shorttitle{}
\shortauthors{Rogers et al.}
\begin{document}

\title{Angular Momentum Transport by Internal Gravity Waves Across Age}

\author[0000-0002-2306-1362]{T.M. Rogers}
\affiliation{School of Mathematics, Statistics and Physics, \\Newcastle University, \\ UK}
\affiliation{Planetary Science Institute,  \\
Tucson, AZ 85721, \\ USA}

\author[0000-0001-5747-8476]{R. P. Ratnasingam}
\affiliation{School of Mathematics, Statistics and Physics, \\Newcastle University, \\ UK}

\begin{abstract}

We present two-dimensional numerical simulations of convection and waves  in a 7M$_{\odot}$ star across stellar ages ranging from zero age to terminal age main sequence. We show that waves efficiently transport angular momentum across the stellar radiative envelope at young ages. However, as the core recedes, leaving behind a ``spike" in the \bvf{} frequency at the convective-radiative interface, the waves are severely attenuated.  This, coupled with the changing stratification throughout the radiation zone, leads to significantly reduced angular momentum transport at later stages on the main sequence.  Indeed the angular momentum transport at mid-main sequence is typically $3-4$ orders of magnitude lower than at zero age, though we expect this to be somewhat mitigated by the chemical mixing also induced by such waves.  We provide measures of the angular momentum transport, both in terms of the divergence of the Reynolds stress and a typical ``wave luminosity".  However, we caution that the angular momentum transport drives shear flows resulting in both slowing and speeding up of radiative interiors.  While the values of Reynolds stress and angular momentum transport are only within the context of these limited simulations, they are not significantly different to those found previously using simpler prescriptions, providing some confidence in their applicability.  

\end{abstract}



\section{Introduction} \label{sec:intro}

Chemical mixing and angular momentum transport in stellar interiors remains a major stumbling block for stellar evolution.  While mixing length theory (MLT) does a reasonable job describing mixing by convection in one-dimensional (1D) stellar evolution models \citep{joycetayar23}, there is no similarly adequate treatment for mixing at convective-radiative boundaries (otherwise known as core boundary mixing, CBM) or within the bulk of radiative regions. The problem is even more dire for angular momentum transport which has the added complication of being a vector and requiring a direction.  Many one-dimensional (1D) stellar evolution models, including the commonly used Modules for Experiments in Stellar Astrophysics (MESA, \citep{mesa_1,mesa_2,mesa_3,mesa_4,mesa_5,mesa_6}), treat chemical mixing and angular momentum transport as a diffusive process.  The Geneva stellar evolution code \citep{geneva08}, the CESTAM code \citep{marques13} and the STAREVOL code \citep{amard19} can treat angular momentum transport through a more consistent advection term, but this is still highly parameterized.    While these simplifications in 1D models are necessary to model stellar evolution, they are very likely not physical, as evidenced by numerous observations \citep{aertsaraa2019}.

These shortcomings have become increasingly apparent as more detailed observations have become available and particularly as asteroseismology has blossomed under multiple space missions such as CoRoT, Kepler and now TESS (see review by \cite{kurtz22}).  These observations have placed a number of constraints on stellar evolution.  Particularly relevant for this paper are the constraints placed on radial differential rotation in both intermediate-mass main-sequence stars and evolved stars.  First, asteroseismic observations of core rotation in evolved stars showed that it did not increase as much as anticipated as they evolved from the main sequence \citep{beck2012,deheuvels2012}.  Similarly, in intermediate-mass main sequence stars differential rotation is lower than anticipated from theory \citep{vanreeth18} and more efficient angular momentum transport is needed to explain slower than expected near core rotation values and the lack of fast rotating cores \citep{ouzzani19}. For the handful of stars where both near-core and surface rotation can be measured, differential rotation appears weak, both in evolved stars \citep{deheuvels2015} and main sequence stars \citep{kurtz2015, saio2015, saio2021,burssens23}. More recently \cite{aerts25} have shown that these core-convecting stars' near core rotation slows as they age.  All told, observations indicate that there is far more angular momentum transport between convective and radiative regions than current evolution models and their prescriptions can account for.  

In most of the literature both chemical mixing and angular momentum transport are described by a one-dimensional diffusion coefficient.  While this may work for some physical mechanisms, this generally has not been demonstrated with more sophisticated numerical simulations.   Here, we report on angular momentum transport by internal gravity waves (IGW) across the main sequence, investigating in particular, the way the receding core and Brunt-Vaisala ``spike"\footnote{We use the term "spike" here to refer to the increased \bvf{} frequency caused by the compositional gradient left behind as the core recedes. This can be clearly seen in the inset of Figure 1.  In general we will use quotation marks to denote colloquially used terms that may not have an exact science definition, but will use the quotes only in the first instance.} left behind at the core-radiative interface, affects the propagation of waves and consequent angular momentum transport.  

We focus on IGW in this study for a number of reasons.  First, the stars we are discussing have convective cores and extended radiative envelopes.  In this configuration, IGW are generated at the convective-radiative boundary and propagate outward toward the stellar surface.  Along the way, their amplitudes increase, because of a drastically decreasing density.  This makes the effects of IGW particularly prominent in these stars, as evident from the wealth of observations \citep{rami18,bowman2019,bowman2020} finding variability that was predicted in \citep{rogers2013,aertsrogers2015}.  Second, IGW act globally and hence are able to transport angular momentum over the bulk of radiative regions as is needed to explain the observations.  Many processes, such as shear instabilities, generally act locally.  Finally, IGW are ubiquitous in these stars, being continuously generated by the convective core.   This work should be considered alongside that of \cite{ashlinmix23}, which investigated chemical mixing as a function of age and mass and \cite{riccardo3} which investigated IGW as a function of age in 3D.

The rest of this Letter is organized as follows: In Section~\ref{sec:numerical_setup} we discuss the numerical setup and describe simulation parameters, in Section~\ref{subsec:results} we discuss the results and in Section~\ref{sec:discussion} we conclude with caveats and discussion about the results in the context of stellar evolution. 

\section{Numerical Setup} \label{sec:numerical_setup}

We solve the full set of hydrodynamic equations in two dimensions representing an equatorial slice of the star \citep{rg05,rogers2013}. The simulations extend from 0.05R$_{*}$ to 0.90R$_{*}$.  To accommodate this large radial extent, while simultaneously having the simulation time reasonable, we use the anelastic approximation, which filters sound waves.  This has the advantage of allowing us to use larger timesteps, covering more integration time overall.  Of course, the down side is that we cannot resolve sound waves, nor can we adequately describe motions close to the sound speed.  In general, the convective motions in these simulations, and in stellar convective cores on the main sequence, are well below the sound speed and sound waves play little role in the dynamics, so this is a decent approximation.  

\begin{eqnarray}\label{eq:cont_eq}
\lefteqn{\nabla \cdot \overline{\rho} \vec{v} = 0} .\\
\label{eq:mom_eq}
\lefteqn{\dxdy{\vec{v}}{t}+(\vec{v}\cdot\nabla)\vec{v}=-\nabla P - Cg\hat{r} +}\nonumber\\
&  &{2(\vec{v}\times\Omega(r))+\overline\nu(\nabla^{2}\vec{v}+\frac{1}{3}\nabla(\nabla\cdot\vec{v}))}\\
\label{eq:energy_eq}
\lefteqn{\dxdy{T}{t}+(\vec{v}\cdot\nabla){T}=-v_{r}(\frac{d\overline{T}}{dr}-(\gamma-1)\overline{T}h_{\rho})+}\nonumber\\
&  & {(\gamma-1)Th_{\rho}v_{r}+\gamma\overline{\kappa}[\nabla^{2}T+(h_{\rho}+h_{\kappa})\dxdy{T}{r}]}
\end{eqnarray}

\begin{figure}
        \centering
        \includegraphics[width=\columnwidth]{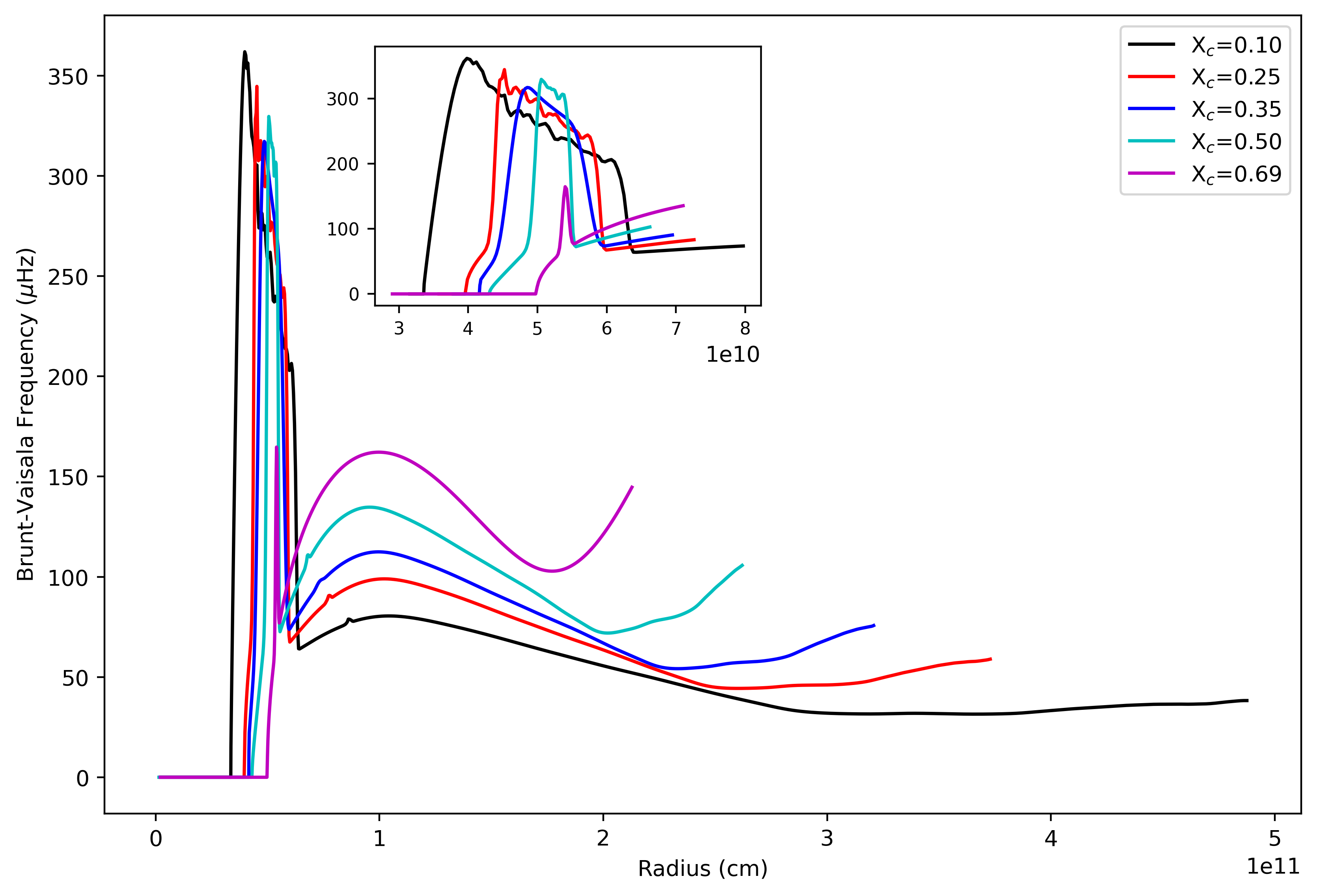}
        \caption{\bvf{} frequency as a function of radius for all ages simulated. The inset shows the zoom-in of the evolution of the near core spike. \label{fig:bvf}}
        \centering
\end{figure}

Equation~\eqref{eq:cont_eq} represents the continuity equation in the anelastic approximation, where $\overline{\rho}$ is the reference state density.  Equation~\eqref{eq:mom_eq} represents the momentum equation, where $\vec{v}$ is the velocity, $P$ is the reduce pressure, C is the co-density \citep{bra95,rg05}, g is gravity, $\Omega\left(r\right)$, is the rotation rate, which is set to a constant, $10^{-6}$ rad s$^{-1}$, within and across these simulations and $\nu$ is the viscous diffusivity set to a constant within and across simulations ($=5\times 10^{12}$ cm$^2$ s$^{-1}$). Equation~\eqref{eq:energy_eq} is the energy equation written as a temperature equation, where  $T$ is the temperature perturbation, while $\overline{T}$ is the reference state temperature.  The vertical velocity is represented as $v_{r}$, $\gamma$ is the ratio of specific heats, $\kappa$ is the thermal diffusivity and $h_{\kappa}$ is its scale height.  $\kappa$ is set to be inversely proportional to the density $\overline{\rho}^{-1/2}$ in all simulations $=5\times 10^{12} (\overline{\rho}(r_0)/\overline{\rho}(r))^{1/2}$ cm$^{2}$ s$^{-1}$, where $r_0$ is the radius of the bottom boundary of the simulated domain to mimic the increasing thermal diffusivity in the radiative zone.   $h_{\rho}$ is the density scale height.  The first term on the right hand side (RHS) of Equation~\eqref{eq:energy_eq} represents the super- or sub-adiabaticity.  It is through this term that we force convective or radiative regions. We note that this formalism does not have boosting\footnote{The term boosting is often used to refer to increasing the stellar luminosity to reduce the thermal relaxation time.} per se, but the superadiabaticity set in all simulations ($10^{-6}$) leads to velocities which are larger than those predicted from MLT. Because this value is set the same in all simulations, all simulations have similar root-mean-square (rms) velocities of around $1-2\times 10^{5}$ cm/s, which is a factor of  $5-10$ larger than MLT velocities (depending on age). This could be interpreted as a boost in the luminosity.  The sub-adiabaticity is taken from the 1D stellar model (see Figure~\ref{fig:bvf}).  The receding convective core leaves behind a spike in the \bvf{} frequency and this is incorporated through the sub-adiabaticity by calculating the first term on the RHS of Equation~\eqref{eq:energy_eq} from the \bvf{} from the 1D reference state model.  

These equations are solved around a reference state model (denoted by overbars in Eqs.\eqref{eq:cont_eq} -- \eqref{eq:energy_eq}), which was calculated using the 1D Modules for Experiments in Astrophysics (MESA) stellar evolution code (version 8848) for a 7M$_{\odot}$ star.  The various models represent different ages, denoted by the hydrogen mass fraction in the core, $X_{\rm c}=0.1,0.25,0.35,0.50~\&~0.69$ (corresponding to $X_{\rm c}/X_{\rm i}=0.14,0.36,0.50,0.71 ~\&~ 0.99$, where $X_{i}$ is the initial hydrogen fraction, set to 0.70).  The inlists used to produce these models are available publicly on \href{https://zenodo.org/record/2596370#.X2sZVHVKg5k}{\textcolor{blue}{zenodo}}. For all the models, stellar metallicity, $Z$, was set to be equal to the solar value of $Z$ = 0.02. The mixing length parameter was set as 1.8, which is a widely-used, solar-calibrated value \citep{Joyce2018}, and the convective overshoot profile was set to exponential, given by $\exp(-2\mathrm{dr}/(f_{ov}H_p)$, dr and $H_p$ are the radial grid size and pressure scale height respectively, and we have set $f_{ov}$ to 0.02 \citep{ehsan2015,claret2016}.

Eqs.\eqref{eq:cont_eq} -- \eqref{eq:energy_eq}) are solved in cylindrical geometry representing an equatorial slice of the star. We use a combined Fourier spectral decomposition in the horizontal dimension and a finite difference discretisation in the vertical.  The spectral method allows for efficiency and accuracy while the finite difference method allows us to be flexible with radial resolution where we need it.  This is particularly relevant in these simulations coupling convective and radiative layers.  All models are run with 2048 horizontal grid points (1440 spectral modes) and 1500 radial zones, distributed such that there are always 500 zones in the convection zone and 1000 zones covering the radiative envelope.  Time-stepping is done using Adams-Bashforth method for the nonlinear terms and Crank-Nicolson for the linear terms.  The code is parallelized using Message Passing Interface (MPI).  Boundary conditions are stress-free and impermeable for velocity at both top and bottom boundaries, while temperature boundary conditions are constant temperature at the top and bottom.  The simulations have run for at least $1.5\times 10^{7}$s. For the total runtime of all the simulations, see Table~\ref{tab:parameters}, where the times are given in units of seconds or convective turnover time, defined as $\tau_{\rm c} = D_{\rm cz}/v_{\mathrm{rms}}$, where $D_{\rm cz}$ is the depth of the convection zone and $v_{\mathrm{rms}}$ is the root-mean-square (rms) velocity over the whole convection zone. The various models run, along with their parameters are shown in Table~\ref{tab:parameters}.  

\begin{table}
\centering
\hskip-1.2cm
\resizebox{1.1\columnwidth}{!}{
\begin{tabular}{cccccc}

\hline
Age ($X_{\rm c}$) & Re  & D$_{cz}$ (10$^{10}$cm,$R_{*}$) & R$_{*}$(10$^{11}$cm) & Time (s, $\tau_{c}$) & $\rho_{in}/\rho_{out}$\\

\hline

0.10& 1005 &3.35,0.07 & 4.78 & $1.6 \times 10^{7}$,71 & 7.02$\times 10^{6}$ \\
0.25& 1185 & 3.95,0.11& 3.76 & $3.2 \times 10^{7}$,121& 1.98$\times 10^{6}$ \\
0.35& 1257 &4.14,0.13 & 3.21 &$1.6 \times 10^{7}$,58 &9.94 $\times 10^{5}$ \\
0.50& 1257 &4.19,0.16 & 2.61 & $1.6\times 10^{7}$,57 & 3.72 $\times 10^{5}$ \\
0.69& 1490 &4.97,0.22 & 2.30 &$7\times10^{7}$,211  &1.11$\times 10^{5}$  \\

\hline
\end{tabular}
}

\caption{Model parameters.  The stellar age, the primary variable in this paper, is represented in core hydrogen fraction.  Re is the Reynolds number in the convective region, calculated using the depth of the convection zone D$_{\rm cz}$ and the rms velocity ($1.5\times 10^{5}$ cm/s).  The depth of the convection zone, D$_{\rm cz}$, is given in both units of $\times 10^{10}$cm, as well as fraction of the simulated radius.  Column 4 shows the simulated radius of the star, R$_{*}$, in $\times 10^{11}$cm, which is always 90\% of the actual radius.   The simulated time is in column 5, represented both in seconds and in convective turnover times.   Finally, the density stratification, represented as density at the inner boundary divided by density at the outer boundary is in column 6.\label{tab:parameters}}

\end{table}

\section{Results} \label{subsec:results}

\begin{figure*}[ht!]
        \centering
        \includegraphics[trim={0.cm 0.cm 0.cm 0.0cm},clip,width=\textwidth]{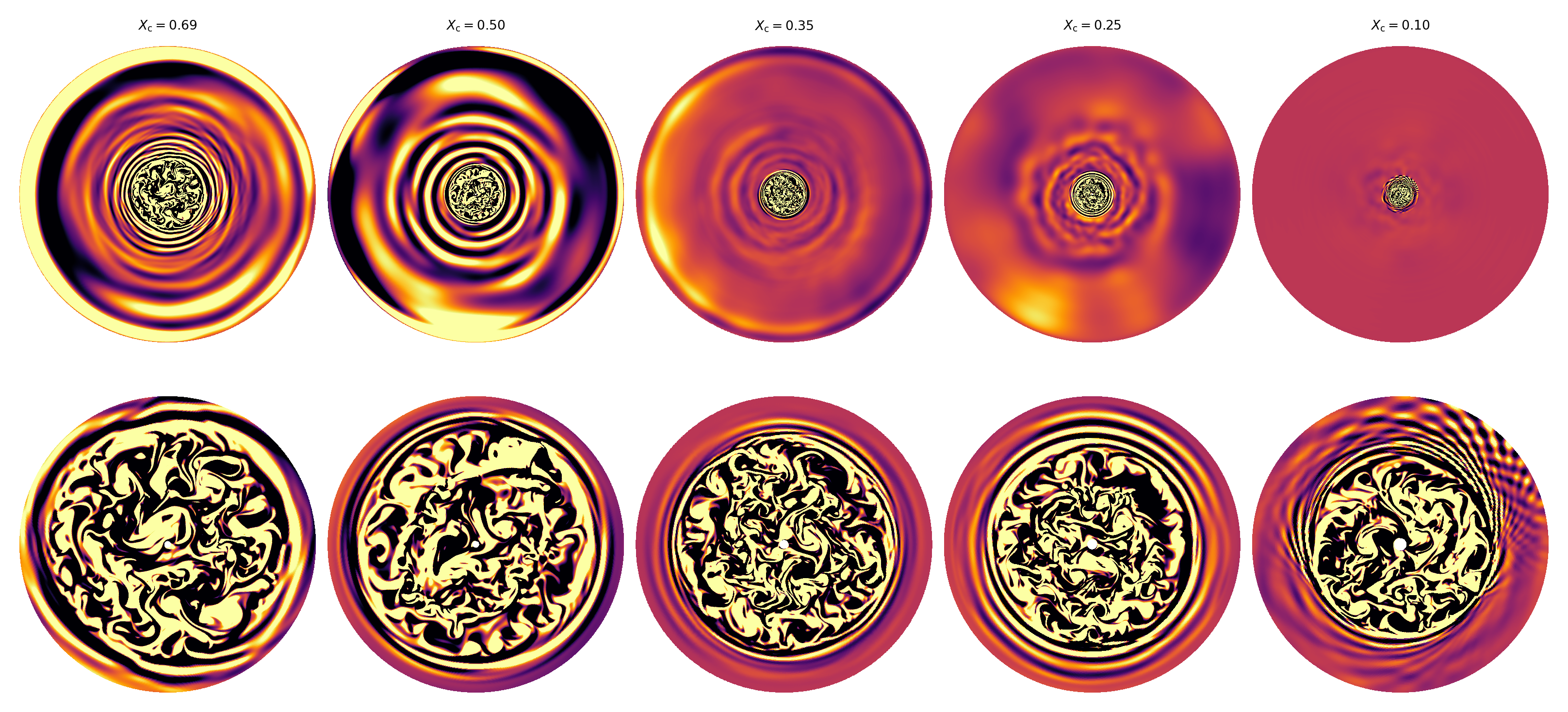}
        \caption{Time snapshots of all simulations, ageing from left to right.  Images are of vorticity with positive (prograde) vorticity white (saturated at 2$\times 10^{-5}$) and negative (retrograde) vorticity black (saturated at -2$\times 10^{-5}$).  Note the varying radial scale as the star ages.  Top row shows full simulated domain, while bottom row shows zoomed in images of the convective core and \bvf{} spike. \label{fig:timesnaps}}
        \centering
\end{figure*}

Figure~\ref{fig:timesnaps} shows time snapshots from all five ages, including zoomed images of the convective cores and \bvf{} spike at the interface.  All ages are shown on the same colour scale in order to highlight the effect of age on the dynamics.  These images demonstrate the basic physics: the convective core generates waves which propagate into the radiative region. As the star ages, the convective core recedes, leaving behind a \bvf{} spike at the convective radiative interface, meanwhile the radiative envelope grows.  At early ages (left) the waves are efficiently generated and propagate to the stellar surface with fairly large amplitudes.  Early in the simulation a variety of waves are present but after many wave crossing times we see mostly large scale, lower frequency waves (evident from the low angle of the wave phase lines in the image).  These large amplitude waves dissipate near the top boundary due to the decreasing density \citep{rogers2013}, driving a large-scale prograde shear flow, seen as positive vorticity at all longitudes near the top.  This is seen at both $X_c=0.69$ and 0.5.  While conservation of vorticity in 2D may contribute to the generation of this shear layer, it is also physical and can be demonstrated to be IGW driven, similar to in \cite{rogers2013}.  At $X_c=0.35$ we can see a marked decrease in the wave amplitudes compared to younger ages and no strong shear flow develops. Simultaneously we can see that the convective core has receded and standing modes are trapped within the \bvf{} spike.  Finally, at late stages, wave amplitudes are decreased even further.  At $X_c=0.1$ we barely see any indication of IGW in the radiative interior (at this color scale), though we see clear indication of wave propagation and trapping in the extended \bvf{} spike.  We also note the increasing depth of the simulation and radiative region leads to reduced numerical resolution, which no doubt contributes to our inability to resolve waves in those simulations.  However, we have run both increased resolution and decreased viscosity in these extended regions and seen virtually no difference.  Moreover, there are multiple physical reasons why wave amplitudes are significantly reduced in aged stars (see Section~\ref{subsec:ang_mom} and \cite{ashlinmix23}).  

\subsection{Reynolds Stress vs. Viscous Stress}\label{subsec:Reynolds}
Quantitatively, the angular momentum transport by waves (any motion) is described by the horizontal average of the momentum equation, given by:

\begin{equation}
\label{eqn:momentumeqn}
\dxdy{\overline{
    U}}{t}+\frac{1}{r\overline{\rho}}\dxdy{~\left(r\langle\overline{\rho} v_{\phi}v_{r}\rangle\right)}{r}=\nu\dxdys{\overline{U}}{r}
\end{equation}
where $v_{\phi}$ is the azimuthal velocity, and $v_{r}$ is the radial velocity and brackets denote a horizontal average.  Eqn.~\ref{eqn:momentumeqn} shows that the mean zonal flow, $\overline{U}$,  is driven by the divergence of the horizontally-averaged Reynolds stress (hereafter referred to as DRS) and is decelerated by viscous dissipation. 

\begin{figure*}[ht!]
        \centering
        \includegraphics[trim={0.cm 0.cm 0.cm 0.0cm},clip,width=\textwidth]{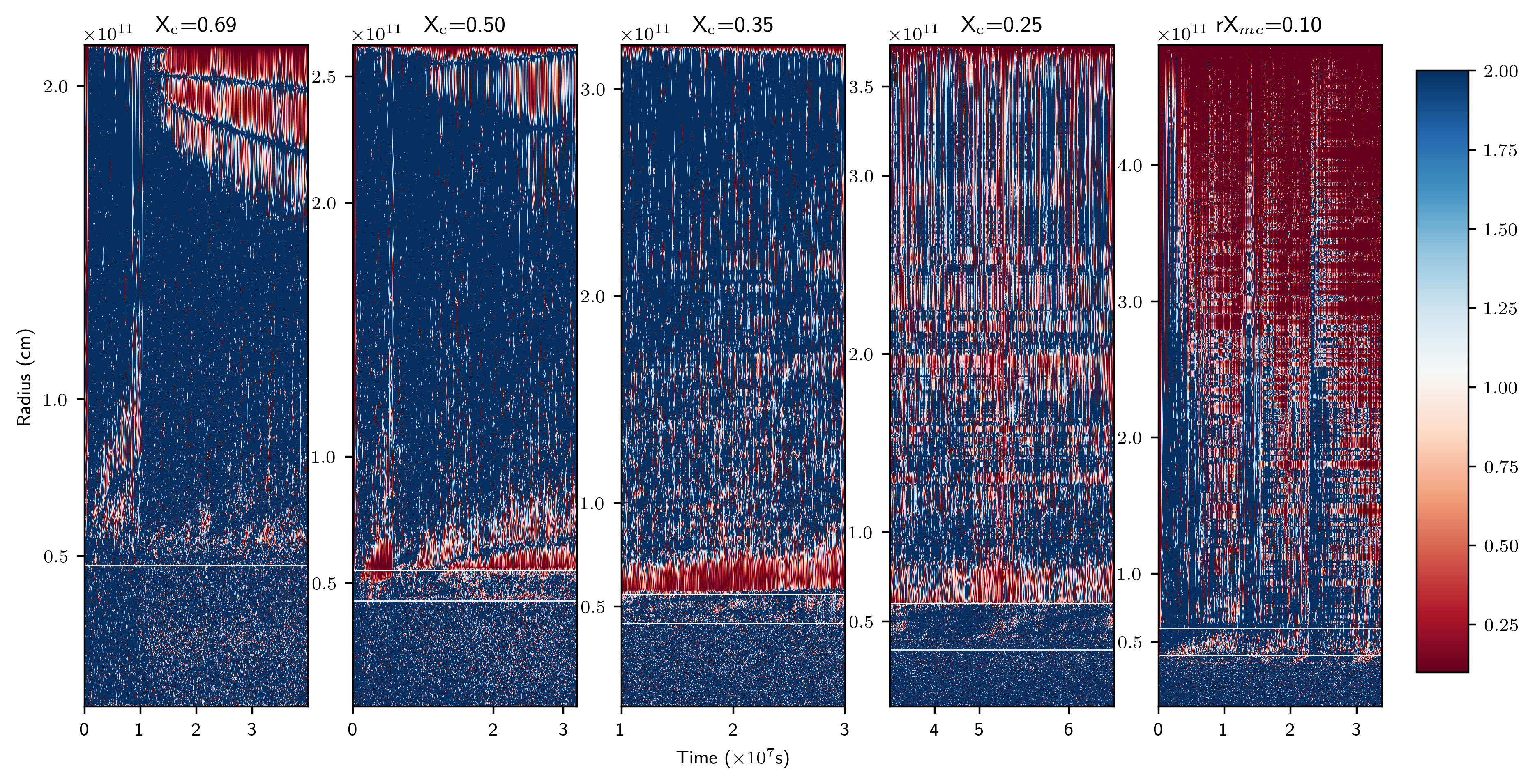}
        \caption{Ratio of the divergence of the Reynolds stress (second term on lhs of ~\eqref{eqn:momentumeqn}) to viscous stress (rhs of ~\eqref{eqn:momentumeqn}).  Blue demonstrates regions that are dominated by Reynolds stress, while red shows regions which are dominated by viscous stress. Horizontal white lines denote the convective radiative interface (for $X_c=0.69$) and the extent of the ~\bvf{} spike for all other ages. Note the varying ordinate axis to reflect the vast expansion in radius across the main sequence lifetime. \label{fig:restressratio}}
        \centering
\end{figure*}

If wave transport is dominant the DRS must be dominant over the viscous term on the right hand side (RHS) of Eq.~\eqref{eqn:momentumeqn}.  In Fig.~\ref{fig:restressratio} we show the ratio of DRS to viscous stress as a function of time and radius, across all ages. There we see that generally, within both the convective and radiative regions, the Reynolds stress is dominant.  Viscous stresses become dominant once large-scale wave-driven shear flows develop at the simulation surface (as seen at $X_c$=0.690 and 0.500) and near the core-radiative envelope (again, due to large scale shear between the regions) and finally, at the latest stages of core-hydrogen burning.  

We note that at $X_{\rm c}=0.1$ the radiative region is entirely dominated by viscous stress. This is both numerical and physical.  At these late stages, because the radiative zone has expanded so significantly, our resolution is not sufficient to resolve this region well. However, the effect is predominantly physical.  The wave amplitudes are severely attenuated by the large \bvf{} spike (Fig.~\ref{fig:bvf}) and the dominant wave frequencies become non-oscillatory in the outer regions of the star, as discussed both in  \cite{ashlinmix23} and \cite{riccardo3}.  This is similarly the case for $X_c=0.25$, though it is less severe.  For these reasons, while we include this simulation for completeness we do so with caution that the results may not be robust, though they are consistent with the trends seen across the age range.  

Another region where viscous stress is dominant is near the simulated stellar surface, particularly at young ages. This is due to the wave-driven large scale shear flow that has developed there, similar to the shear flows that drive the Quasi-Biennial-Oscillation \citep{ba01,Plumb77} and those seen in experiments such as \cite{pm78} or the more recent simulations from \cite{couston18}. In these simulations, once the flows develop, the dominant contribution to the momentum transport is viscous stress, which acts on a very long timescale, so a reversal is not observed.  Given the depth of the flow and the viscosity, we would expect this to take on a timescale of order a few times $10^{8}$s. While this is reduced due to the action of the waves, it is still much longer than we have run these simulations. Moreover, if we were to reduce our viscous diffusion, making the simulation closer to stellar conditions, this timescale would increase even further.  

\begin{figure}[ht!]
        \centering
        \includegraphics[trim={1.cm 1.cm 0.cm 1.7cm},clip,width=\columnwidth]{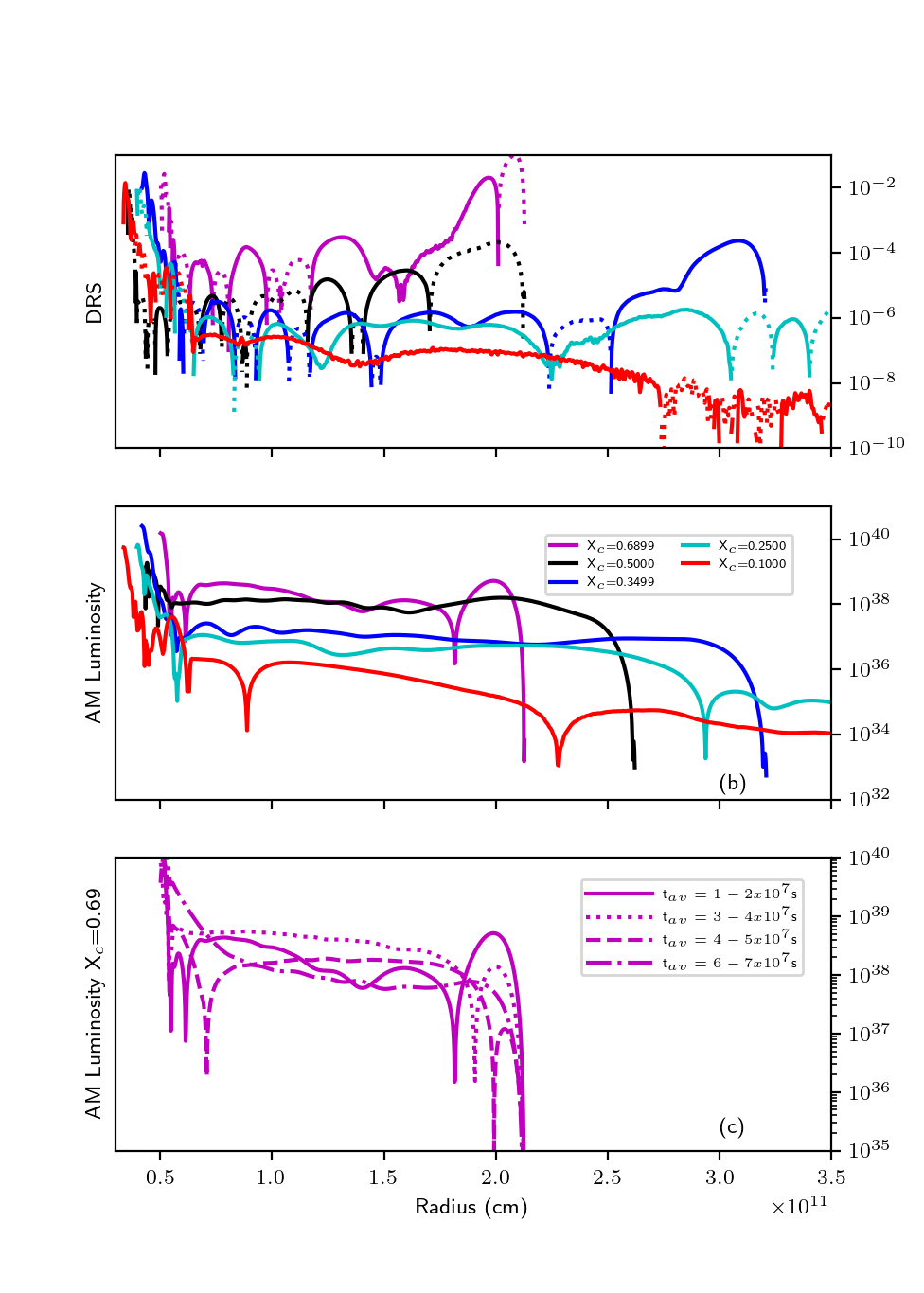}
        \caption{(a) Time-averaged Reynolds stress (second term in ~\eqref{eqn:momentumeqn}) as a function of radius for all ages.  Averaged over the last 5$\times 10^{6}$s for each model, i.e. from $1\times 10^{7}-1.5\times 10^{7}$s. (b) Absolute value of the angular momentum luminosity, as defined in \cite{TC03}, (c) Time-averaged Reynolds stress for $X_c=0.69$ for the same time interval shown in (a), solid line, and for $3-4\times 10^7$s (dotted), $4-5\times 10^{7}$s (dashed) and $5-6\times 10^{7}$s.  All quantities are in cgs units. The data for (a) and (b) is available on \href{https://doi.org/10.5281/zenodo.15044247}{\textcolor{blue}{zenodo}}.  \label{fig:restressage}}
        \centering
\end{figure}

Finally, we highlight that the Reynolds stress is dominant in the \bvf{} spike (denoted by horizontal white lines for all stars), for nearly all ages.   From Fig.~\ref{fig:restressage}, we can conclude that at most times and locations, apart from the exceptions discussed above, the Reynolds stress is dominant.  This should ensure that the angular momentum transport we observe in the radiation zone is indeed due to IGW.

\subsection{Angular Momentum Transport by Waves}\label{subsec:ang_mom}
In order to investigate the impact of IGW on angular momentum transport in stars, we show the long timescale average (over 10$^{7}$s) of the absolute value of the DRS in Fig.~\ref{fig:restressage}a, for all stellar ages simulated.    There are many important features to highlight in that plot. First, at a given age, the DRS varies substantially with radius.  For example, the Reynolds stress amplitude varies by three orders of magnitude for $X_c =0.69$, initially dropping with radius and then increasing again near the surface.  The variation is less severe with age, but still present.  The initial drop observed at all ages is due to the rapid dissipation of low frequency waves outside the convection zone.  The increase near the stellar surface is due to the amplification of wave amplitude from the density stratification. This is enhanced by the mean-flow those waves generated at $X_c =0.69$ and $0.50$. Second,  the DRS changes sign throughout the bulk of the radiative interior at all ages, with prograde flows demonstrated by solid sections and  retrograde flows demonstrated by the dotted sections of the lines.  This is due to the ``anti-diffusive"\footnote{IGW drive shear flows, due to the frequency dependence of damping.} nature of IGW angular momentum transport \citep{Plumb77,pm78}.  Finally, and perhaps most importantly, the Reynolds stress varies by approximately three orders of magnitude across stellar (main-sequence) age. This reduction across age is due to numerous effects, which were also discussed in relation to chemical mixing \citep{ashlinmix23}.  In older stars, lower frequencies can propagate through the \bvf{} spike, but are more rapidly damped in the outer layers of the radiative envelope, meanwhile higher frequencies are trapped and hence, cannot contribute to the angular momentum transport.  Moreover, at later stages of evolution the turning point (where the ratio of the oscillatory term to the density term in the wave equation becomes less than unity \citep{rathish2020,ashlinmix23,riccardo3}),  moves inwards in radius.  Hence, waves lose their oscillatory nature and become virtually evanescent.  All of these effects act to reduce wave amplitudes at later stages of evolution, leading to reduced angular momentum transport. However, as shown in \cite{ashlinmix23} IGW also cause chemical mixing which would erode the height and extent of the \bvf{} spike and hence, the attenuation of waves.

Most calculations of the impact of IGW on stellar interiors use the flux formulation laid out in \citep{he63,bre66,ktz99}, where the DRS is replaced by the divergence of a ``flux" of individual waves (sometimes referred to as the ``wave luminosity", once integrated over frequency and wavelength). The individual waves are thermally dissipated and their collective action is approximated by summing over the (relevant) individual waves, e.g.~\citealt{TC03,fuller2014}.  Here, we calculate the luminosity directly from the horizontally averaged Reynolds stress in order to make contact with previous work and so that these simulation results can be of use to 1D modeling efforts. We define the wave luminosity as in \cite{TC03} such that $L(r) \approx 8 \pi r^2  \langle \overline{\rho} r v_r v_{\phi}\rangle/3$, which is shown\footnote{We note that this equation is for a sphere and these are obviously 2D simulations, we include r$^2$ instead of r for comparison and simplicity in implementation. } in Fig.~\ref{fig:restressage}b (neglecting the geometrical factor of $8\pi/3$). There we similarly see an approximately three order of magnitude variation from $X_{\rm c}=0.69$ to $X_{\rm c}=0.1$ in wave luminosity.  Unlike the DRS, the luminosity is rather flat with radius, except at the oldest ages, where the trend is decreasing, again, likely due to the reduction of the oscillatory term in the wave equation.  

To our knowledge, the only published wave luminosities, integrated over wavelength, frequency and time, attempting to account for both prograde and retrograde waves, are those in \cite{TC03}, for solar-type stars. Those values range from $10^{33}-10^{36}$ erg, depending on the stellar mass.   One would expect IGW luminosities to be larger in more massive stars with core convection, if the simple prescription from \cite{GoldreichKumar1990}, $L_w\sim \mathcal{M} L_*$, where $\mathcal{M}$ is the Mach number, $L_w$ is the wave luminosity and $L_*$ is the stellar luminosity, holds.  In this case one would expect the wave luminosity to increase by $\sim 300-1000$ for this 7M$_{\odot}$ star compared to the Sun, due to the increased luminosity and the increased Mach number (luminosity increases by $\sim$7$^3$ and Mach number increases by $\sim$3).  This is similar to what is recovered here compared to the results of \cite{TC03}.  We could also calculate the timescale over which waves could change the angular velocity of the radiative envelope, as in \cite{fuller2014} and \cite{riccardo3} and find timescales ranging from 10$^{9}$s for X$_{c}$=0.69 to 10$^{13}$s for X$_{c}$=0.10.  Even at the oldest ages and lowest angular momentum transport, these timescales are still shorter than the main sequence age, if integrated across age.\footnote{We note that there was a significant unit error in \cite{riccardo3} in which the timescale is quoted as 10$^8$ years rather than \it{seconds}.}  Again, this efficiency is similar to that quoted in e.g. \cite{fuller2014}.  The similarity of the overall results indicates some level of robustness,  given both the shortcomings of these simulations (non-stellar parameters, reduced geometry) and the shortcomings of the theoretical work (linear, non-interacting waves, prescribed filtering, limited spectral contribution). Similar to \cite{ashlinmix23} we find that the dominant frequencies in the angular momentum transport are large scale and in the frequency range of $6-12\mu$Hz.  These waves represent a balance between generation amplitude and weak thermal damping.  Therefore, these are likely the waves that should be included in the flux formulation in stellar evolution codes, where non-diffusive transport is possible \citep{geneva08,marques13}.  

Though 10$^7$s is a long timescale for these simulations, it is a small amount of time in an evolutionary sense and it is wise to wonder if this timescale is sufficient to extrapolate to stellar lifetimes.  While we cannot know for certain without running our simulations over an evolutionary time, which is not possible, we can compare the values retrieved on this timescale to different timescales for $X_c=0.69$, which was run significantly longer, this is shown in Fig.~\ref{fig:restressage}c. There, we see that the Reynolds stress amplitude, structure and variation with radius is largely similar, with amplitudes varying by about $5-10$ depending on the radial location.  While this is not completely stable, it is likely reflective of the time dependent nature of stellar interiors generally and averaging of prograde and retrograde flows in time.  

\subsection{Rotation Profiles}\label{subsec:near_core_rot}
\begin{figure}
        \centering
        \includegraphics[trim={0.cm 0.2cm 0.cm 0.cm},clip,width=\columnwidth]{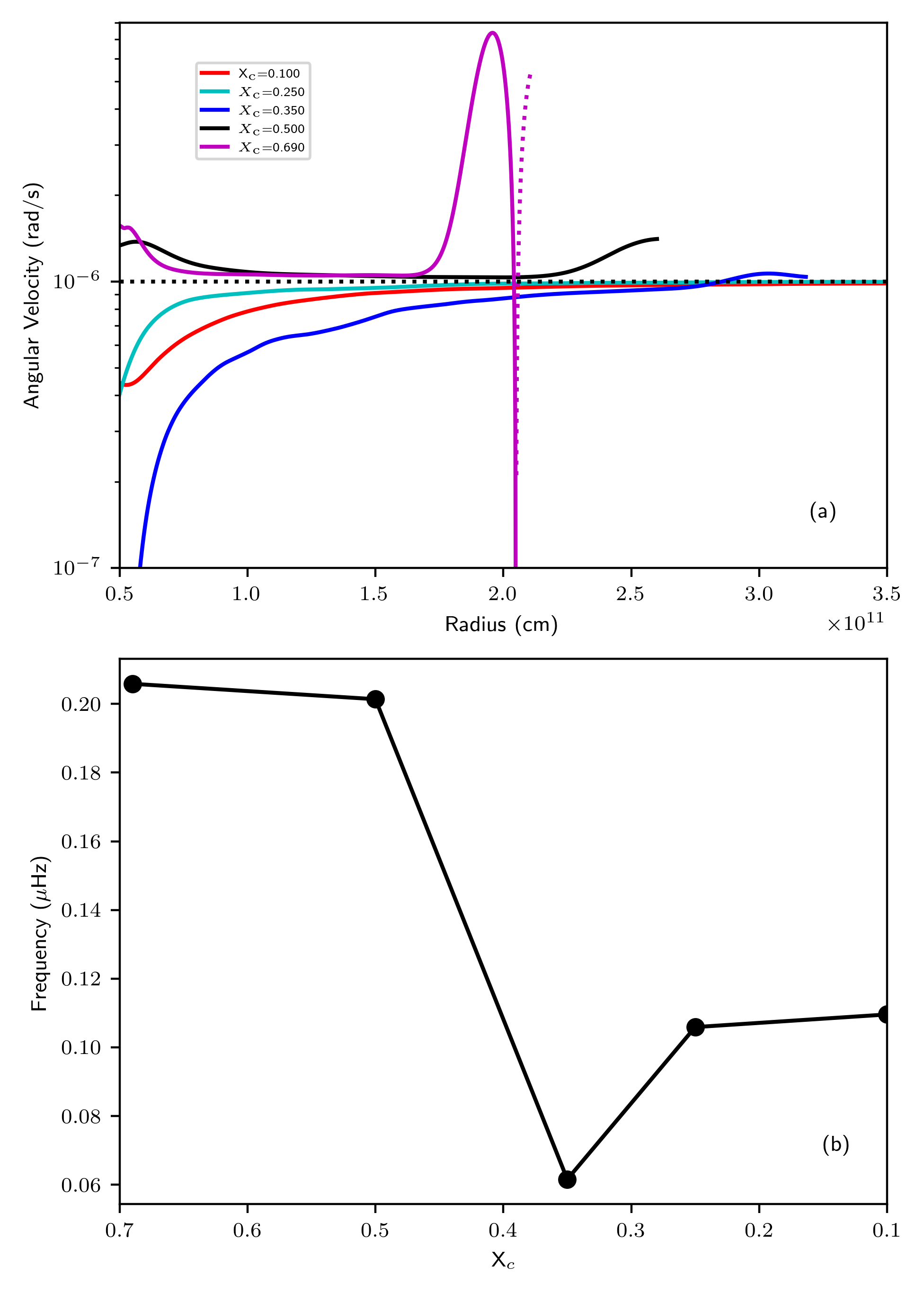}
        \caption{(a) Time and longitude averaged angular velocity.  Time-averaging is done from $1-1.5\times 10^{7}$s for all models, with different colors representing different ages as seen in the legend. The dotted part of the X$_c$=0.69 line represents a negative angular velocity. (b) Rotational frequency as a function of age, averaged over the same time period and calculated using a \bvf{}-weighted average as in \cite{ouzzani19}.}\label{fig:rotation-age}
        \centering
\end{figure}

Ultimately, we are concerned with the rotation profiles that result from IGW transport. We show the rotation profiles in the bulk of the radiation zone for each age in Fig.~\ref{fig:rotation-age}a, averaged over 1-1.5$\times 10^{7}$s.  There we see that, in general, IGW speed up the radiation zone (from the initial rotation, indicated by the horizontal dotted line) at younger ages, and {\it slow down} the radiation zone at later stages of the main sequence.  Shear flows, driven by IGW, are clearly seen at the surface at younger ages, even as late as X$_{\rm c}$=0.35 (note that we do not show the surface regions of the older models for clarity, no surface shear develops in these models).  We note that such angular momentum transport would exasperate core-envelope decoupling at later stages of main sequence evolution.

To look at an integrated view of the angular velocity in the radiation zone we show the \bvf{} weighted average of the rotation rate, as defined in \cite{ouzzani19}, for each age in Fig.~\ref{fig:rotation-age}b.     Given that the Reynolds stress is dominant in most of the models this is another indicator of how angular momentum transport by IGW varies across age, and in particular in the near-core region, as this is where the average is most sensitive to.  We see that the rotation rate generally decreases with age.  This is broadly consistent with the recent results from \cite{aerts25}, in the sense that overall the "near-core" rotation slows across age and that the slowing is approximately a factor of two.  However, there are several important caveats.  First, we start with substantially lower rotation rates than expected in that sample. Second, that sample is dominated by F stars, with tenuous surface convection zones which possibly contribute to the near-core rotation rate. Finally, and most importantly, our simulations do not conserve angular momentum across the different ages. Therefore, we have not accounted for the differential rotation that would naturally develop between the core and envelope.  Since we do not account for the slow down of the radiation zone due to expansion, what is measured in Figure 5 is a reduction in angular velocity purely due to IGW transport.

\section{Discussion}\label{sec:discussion}
The results presented here lack some possibly important physics.  First, we note that these simulations were all performed at a fairly slow rotation which may not be representative of real stars and a forthcoming paper will investigate the role of rotation on the angular momentum transport.  Second, as mentioned above, older stars start to rotate differentially as the core contracts and spins up, while the surface expands and slows down, which is not accounted for in these simulations. Moreover, older stars likely develop surface convection zones and the interaction of surface generated waves, with core generated waves, could greatly impact the angular momentum transport and mixing due to waves at later stages. Finally, as shown in \cite{ashlinmix23} the IGW will induce chemical mixing, which would reduce the effect of the \bvf{} spike.  Therefore, due to all of the effects of differential rotation, chemical mixing and surface convection zones we expect the reduction across age to be less severe than seen here. 

The simulations themselves are also limited in a variety of ways.  First, they are 2D which could impact the dissipation of waves and the generation of shear flows.  Second, our (explicit) viscous and thermal diffusion coefficients are much larger than a real star, leading to enhanced dissipation of the waves and reduced turbulence in the convection zone.   On the other hand, our enhanced convective velocities counteract, at some level, these enhanced diffusivities, which allows our vertical wave displacements to be large compared to a grid cell. 

We note that these shortcomings are true for most hydrodynamic simulations. Still, our simulations have some advantages over many others, such as being able to run longer for less computer time (due to the anelastic approximation) and our use of explicit viscosities, while large, means we know fundamental parameters, such as Reynolds number.  Moreover, while the geometry is limited, it is more realistic than Cartesian coordinates because it allows more accurate representation of wave propagation and standing modes in a star.  These approximations have allowed us to run simulations across a variety of masses \citep{rogers2013,philipp3dpaper,rathish2020,ashlinmix23} and ages \citep{ashlinmix23,riccardo3}. Still, the limitations must be recognized and the results presented here should be used with an understanding of those limitations. 

Despite the above shortcomings, we have made the following robust conclusions: 1) angular momentum transport by waves is efficient at early stages of the main sequence, but drops as the star ages and 2) the amplitude of the angular momentum luminosity is larger than in previous studies, but consistent with expectations from simple prescriptions for wave luminosity and is fairly flat across radius,though caution should be exercised given that sometimes this wave luminosity is effectively negative.

% In the bibliography the format for data or code follows this format: \\

% \noindent author year, title, version, publisher, prefix:identifier\\

% \citet{2015ApJ...805...23C} provides a example of how the citation in the
% article references the external code at
% \doi{10.5281/zenodo.15991}.  Unfortunately, bibtex does
% not have specific bibtex entries for these types of references so the
% ``@misc'' type should be used.  The Repository tutorial explains how to
% code the ``@misc'' type correctly.  The most recent aasjournal.bst file,
% available with \aastex\ v6, will output bibtex ``@misc'' type properly.

%% Also note that the akcnowlodgment environment does not support long amounts of text. If you have a lot of people and institutions to acknowledge, do not use this command. Instead, create a new \section{Acknowledgments}.
\begin{acknowledgments}
We acknowledge support from STFC grant ST/W001020/1. We thank C. Aerts, A. Varghese, P. Edelmann and D. Bowman  for helpful conversations when carrying out this work and preparing the manuscript.  We also thank an anonymous referee for helpful comments which improved the presentation and text.
\end{acknowledgments}

\bibliographystyle{aasjournal}
\bibliography{igwcl}{}

\newcommand{\noop}[1]{}
\begin{thebibliography}{}
\expandafter\ifx\csname natexlab\endcsname\relax\def\natexlab#1{#1}\fi
\providecommand{\url}[1]{\href{#1}{#1}}
\providecommand{\dodoi}[1]{doi:~\href{http://doi.org/#1}{\nolinkurl{#1}}}
\providecommand{\doeprint}[1]{\href{http://ascl.net/#1}{\nolinkurl{http://ascl.net/#1}}}
\providecommand{\doarXiv}[1]{\href{https://arxiv.org/abs/#1}{\nolinkurl{https://arxiv.org/abs/#1}}}

\bibitem[{{Aerts} {et~al.}(2019){Aerts}, {Mathis}, \& {Rogers}}]{aertsaraa2019}
{Aerts}, C., {Mathis}, S., \& {Rogers}, T.~M. 2019, \araa, 57, 35, \dodoi{10.1146/annurev-astro-091918-104359}

\bibitem[{{Aerts} \& {Rogers}(2015)}]{aertsrogers2015}
{Aerts}, C., \& {Rogers}, T.~M. 2015, \apjl, 806, L33, \dodoi{10.1088/2041-8205/806/2/L33}

\bibitem[{{Aerts} {et~al.}(2025){Aerts}, {Van Reeth}, {Mombarg}, \& {Hey}}]{aerts25}
{Aerts}, C., {Van Reeth}, T., {Mombarg}, J. S.~G., \& {Hey}, D. 2025, arXiv e-prints, arXiv:2502.17692, \dodoi{10.48550/arXiv.2502.17692}

\bibitem[{{Amard} {et~al.}(2019){Amard}, {Palacios}, {Charbonnel}, {Gallet}, {Georgy}, {Lagarde}, \& {Siess}}]{amard19}
{Amard}, L., {Palacios}, A., {Charbonnel}, C., {et~al.} 2019, \aap, 631, A77, \dodoi{10.1051/0004-6361/201935160}

\bibitem[{Baldwin {et~al.}(2001)Baldwin, Gray, Dunkerton, Hamilton, Hayes, Randel, Holton, Alexander, Hirota, Horinouchi, Jones, Kinnersley, Marquardt, Sato, \& Takahashi}]{ba01}
Baldwin, M., Gray, L., Dunkerton, T., {et~al.} 2001, Reviews of Geophysics, 39, 179

\bibitem[{{Beck} {et~al.}(2012){Beck}, {Montalban}, {Kallinger}, {De Ridder}, {Aerts}, {Garc{\'{\i}}a}, {Hekker}, {Dupret}, {Mosser}, {Eggenberger}, {Stello}, {Elsworth}, {Frandsen}, {Carrier}, {Hillen}, {Gruberbauer}, {Christensen-Dalsgaard}, {Miglio}, {Valentini}, {Bedding}, {Kjeldsen}, {Girouard}, {Hall}, \& {Ibrahim}}]{beck2012}
{Beck}, P.~G., {Montalban}, J., {Kallinger}, T., {et~al.} 2012, \nat, 481, 55, \dodoi{10.1038/nature10612}

\bibitem[{{Bowman} {et~al.}(2020){Bowman}, {Burssens}, {Sim{\'o}n-D{\'\i}az}, {Edelmann}, {Rogers}, {Horst}, {R{\"o}pke}, \& {Aerts}}]{bowman2020}
{Bowman}, D.~M., {Burssens}, S., {Sim{\'o}n-D{\'\i}az}, S., {et~al.} 2020, \aap, 640, A36, \dodoi{10.1051/0004-6361/202038224}

\bibitem[{Bowman {et~al.}(2019)Bowman, Burssens, Pedersen, Johnston, Aerts, Buysschaert, Michielsen, Tkachenko, Rogers, Edelmann, Ratnasingam, Sim\'on-D\'{\i}az, Castro, Moravveji, Pope, White, \& Cat}]{bowman2019}
Bowman, D.~M., Burssens, S., Pedersen, M., {et~al.} 2019, Nature Astronomy, 3, \dodoi{10.1038/s41550-019-0768-1}

\bibitem[{Braginsky \& Roberts(1995)}]{bra95}
Braginsky, S., \& Roberts, P. 1995, GAFD, 79, 1

\bibitem[{Bretherton(1966)}]{bre66}
Bretherton, F. 1966, Quarterly Journal of the Royal Meteorolgic Society, 92, 466

\bibitem[{{Burssens} {et~al.}(2023){Burssens}, {Bowman}, {Michielsen}, {Sim{\'o}n-D{\'\i}az}, {Aerts}, {Vanlaer}, {Banyard}, {Nardetto}, {Townsend}, {Handler}, {Mombarg}, {Vanderspek}, \& {Ricker}}]{burssens23}
{Burssens}, S., {Bowman}, D.~M., {Michielsen}, M., {et~al.} 2023, Nature Astronomy, 7, 913, \dodoi{10.1038/s41550-023-01978-y}

\bibitem[{{Claret, A.} \& {Torres, G.}(2016)}]{claret2016}
{Claret, A.}, \& {Torres, G.} 2016, A\&A, 592, A15, \dodoi{10.1051/0004-6361/201628779}

\bibitem[{{Couston} {et~al.}(2018){Couston}, {Lecoanet}, {Favier}, \& {Le Bars}}]{couston18}
{Couston}, L.-A., {Lecoanet}, D., {Favier}, B., \& {Le Bars}, M. 2018, \prl, 120, 244505, \dodoi{10.1103/PhysRevLett.120.244505}

\bibitem[{{Deheuvels} {et~al.}(2015){Deheuvels}, {Ballot}, {Beck}, {Mosser}, {{\O}stensen}, {Garc{\'\i}a}, \& {Goupil}}]{deheuvels2015}
{Deheuvels}, S., {Ballot}, J., {Beck}, P.~G., {et~al.} 2015, \aap, 580, A96, \dodoi{10.1051/0004-6361/201526449}

\bibitem[{Deheuvels {et~al.}(2012)Deheuvels, Garcia, Chaplin, Basu, Antia, Appourchaux, Benomar, Davies, Elsworth, Gizon, Goupil, Reese, Regulo, Schou, Stahn, Casagrande, Christensen-Dalsgaard, Fischer, Hekker, Kjeldsen, Mathur, Mosser, Pinsonneault, Valenti, Christiansen, Kinemuchi, \& Mullally}]{deheuvels2012}
Deheuvels, S., Garcia, R.~A., Chaplin, W.~J., {et~al.} 2012, \apj, 756, 19.
\newblock \url{http://stacks.iop.org/0004-637X/756/i=1/a=19}

\bibitem[{Edelmann {et~al.}(2019)Edelmann, Ratnasingam, Pedersen, Bowman, Prat, \& Rogers}]{philipp3dpaper}
Edelmann, P., Ratnasingam, R., Pedersen, M., {et~al.} 2019, \apj, 876, 4, \dodoi{10.3847/1538-4357/ab12df}

\bibitem[{{Eggenberger} {et~al.}(2008){Eggenberger}, {Meynet}, {Maeder}, {Hirschi}, {Charbonnel}, {Talon}, \& {Ekstr{\"o}m}}]{geneva08}
{Eggenberger}, P., {Meynet}, G., {Maeder}, A., {et~al.} 2008, \apss, 316, 43, \dodoi{10.1007/s10509-007-9511-y}

\bibitem[{{Fuller} {et~al.}(2014){Fuller}, {Lecoanet}, {Cantiello}, \& {Brown}}]{fuller2014}
{Fuller}, J., {Lecoanet}, D., {Cantiello}, M., \& {Brown}, B. 2014, \apj, 796, 17, \dodoi{10.1088/0004-637X/796/1/17}

\bibitem[{{Goldreich} \& {Kumar}(1990)}]{GoldreichKumar1990}
{Goldreich}, P., \& {Kumar}, P. 1990, \apj, 363, 694, \dodoi{10.1086/169376}

\bibitem[{Herring(1963)}]{he63}
Herring, J. 1963, Journal of the Atmospheric Sciences, 20, 325

\bibitem[{{Jermyn} {et~al.}(2023){Jermyn}, {Bauer}, {Schwab}, {Farmer}, {Ball}, {Bellinger}, {Dotter}, {Joyce}, {Marchant}, {Mombarg}, {Wolf}, {Sunny Wong}, {Cinquegrana}, {Farrell}, {Smolec}, {Thoul}, {Cantiello}, {Herwig}, {Toloza}, {Bildsten}, {Townsend}, \& {Timmes}}]{mesa_6}
{Jermyn}, A.~S., {Bauer}, E.~B., {Schwab}, J., {et~al.} 2023, \apjs, 265, 15, \dodoi{10.3847/1538-4365/acae8d}

\bibitem[{{Joyce} \& {Chaboyer}(2018)}]{Joyce2018}
{Joyce}, M., \& {Chaboyer}, B. 2018, \apj, 856, 10, \dodoi{10.3847/1538-4357/aab200}

\bibitem[{{Joyce} \& {Tayar}(2023)}]{joycetayar23}
{Joyce}, M., \& {Tayar}, J. 2023, Galaxies, 11, 75, \dodoi{10.3390/galaxies11030075}

\bibitem[{Kumar {et~al.}(1999)Kumar, Talon, \& Zahn}]{ktz99}
Kumar, P., Talon, S., \& Zahn, J. 1999, ApJ, 520, 859

\bibitem[{{Kurtz}(2022)}]{kurtz22}
{Kurtz}, D.~W. 2022, \araa, 60, 31, \dodoi{10.1146/annurev-astro-052920-094232}

\bibitem[{{Kurtz} {et~al.}(2015){Kurtz}, {Saio}, {Takata}, {Shibahashi}, {Murphy}, \& {Sekii}}]{kurtz2015}
{Kurtz}, D.~W., {Saio}, H., {Takata}, M., {et~al.} 2015, in European Physical Journal Web of Conferences, Vol. 101, European Physical Journal Web of Conferences, 01007, \dodoi{10.1051/epjconf/201510101007}

\bibitem[{{Marques} {et~al.}(2013){Marques}, {Goupil}, {Lebreton}, {Talon}, {Palacios}, {Belkacem}, {Ouazzani}, {Mosser}, {Moya}, {Morel}, {Pichon}, {Mathis}, {Zahn}, {Turck-Chi{\`e}ze}, \& {Nghiem}}]{marques13}
{Marques}, J.~P., {Goupil}, M.~J., {Lebreton}, Y., {et~al.} 2013, \aap, 549, A74, \dodoi{10.1051/0004-6361/201220211}

\bibitem[{{Moravveji, E.} {et~al.}(2015){Moravveji, E.}, {Aerts, C.}, {Pápics, P. I.}, {Triana, S. A.}, \& {Vandoren, B.}}]{ehsan2015}
{Moravveji, E.}, {Aerts, C.}, {Pápics, P. I.}, {Triana, S. A.}, \& {Vandoren, B.} 2015, A\&A, 580, A27, \dodoi{10.1051/0004-6361/201425290}

\bibitem[{{Ouazzani} {et~al.}(2019){Ouazzani}, {Marques}, {Goupil}, {Christophe}, {Antoci}, {Salmon}, \& {Ballot}}]{ouzzani19}
{Ouazzani}, R.~M., {Marques}, J.~P., {Goupil}, M.~J., {et~al.} 2019, \aap, 626, A121, \dodoi{10.1051/0004-6361/201832607}

\bibitem[{{Paxton} {et~al.}(2011){Paxton}, {Bildsten}, {Dotter}, {Herwig}, {Lesaffre}, \& {Timmes}}]{mesa_1}
{Paxton}, B., {Bildsten}, L., {Dotter}, A., {et~al.} 2011, \apjs, 192, 3, \dodoi{10.1088/0067-0049/192/1/3}

\bibitem[{{Paxton} {et~al.}(2013){Paxton}, {Cantiello}, {Arras}, {Bildsten}, {Brown}, {Dotter}, {Mankovich}, {Montgomery}, {Stello}, {Timmes}, \& {Townsend}}]{mesa_2}
{Paxton}, B., {Cantiello}, M., {Arras}, P., {et~al.} 2013, \apjs, 208, 4, \dodoi{10.1088/0067-0049/208/1/4}

\bibitem[{Paxton {et~al.}(2015)Paxton, Marchant, Schwab, Bauer, Bildsten, Cantiello, Dessart, Farmer, Hu, Langer, Townsend, Townsley, \& Timmes}]{mesa_3}
Paxton, B., Marchant, P., Schwab, J., {et~al.} 2015, \apjs, 220, 15, \dodoi{10.1088/0067-0049/220/1/15}

\bibitem[{{Paxton} {et~al.}(2018){Paxton}, {Schwab}, {Bauer}, {Bildsten}, {Blinnikov}, {Duffell}, {Farmer}, {Goldberg}, {Marchant}, {Sorokina}, {Thoul}, {Townsend}, \& {Timmes}}]{mesa_4}
{Paxton}, B., {Schwab}, J., {Bauer}, E.~B., {et~al.} 2018, \apjs, 234, 34, \dodoi{10.3847/1538-4365/aaa5a8}

\bibitem[{{Paxton} {et~al.}(2019){Paxton}, {Smolec}, {Schwab}, {Gautschy}, {Bildsten}, {Cantiello}, {Dotter}, {Farmer}, {Goldberg}, {Jermyn}, {Kanbur}, {Marchant}, {Thoul}, {Townsend}, {Wolf}, {Zhang}, \& {Timmes}}]{mesa_5}
{Paxton}, B., {Smolec}, R., {Schwab}, J., {et~al.} 2019, \apjs, 243, 10, \dodoi{10.3847/1538-4365/ab2241}

\bibitem[{Plumb(1977)}]{Plumb77}
Plumb, R. 1977, J. Atmos. Science, 34, 1847

\bibitem[{Plumb \& McEwan(1978)}]{pm78}
Plumb, R., \& McEwan, A. 1978, J. Atmos. Science, 35, 1827

\bibitem[{{Ramiaramanantsoa} {et~al.}(2018){Ramiaramanantsoa}, {Ratnasingam}, {Shenar}, {Moffat}, {Rogers}, {Popowicz}, {Kuschnig}, {Pigulski}, {Handler}, {Wade}, {Zwintz}, \& {Weiss}}]{rami18}
{Ramiaramanantsoa}, T., {Ratnasingam}, R., {Shenar}, T., {et~al.} 2018, \mnras, 480, 972, \dodoi{10.1093/mnras/sty1897}

\bibitem[{Ratnasingam {et~al.}(2020)Ratnasingam, Edelmann, \& Rogers}]{rathish2020}
Ratnasingam, R.~P., Edelmann, P. V.~F., \& Rogers, T.~M. 2020, \mnras, 497, 4231, \dodoi{10.1093/mnras/staa2296}

\bibitem[{Rogers \& Glatzmaier(2005)}]{rg05}
Rogers, T., \& Glatzmaier, G. 2005, MNRAS, 364, 1135

\bibitem[{{Rogers} {et~al.}(2013){Rogers}, {Lin}, {McElwaine}, \& {Lau}}]{rogers2013}
{Rogers}, T.~M., {Lin}, D.~N.~C., {McElwaine}, J.~N., \& {Lau}, H.~H.~B. 2013, \apj, 772, 21, \dodoi{10.1088/0004-637X/772/1/21}

\bibitem[{{Saio} {et~al.}(2015){Saio}, {Kurtz}, {Takata}, {Shibahashi}, {Murphy}, {Sekii}, \& {Bedding}}]{saio2015}
{Saio}, H., {Kurtz}, D.~W., {Takata}, M., {et~al.} 2015, \mnras, 447, 3264, \dodoi{10.1093/mnras/stu2696}

\bibitem[{{Saio} {et~al.}(2021){Saio}, {Takata}, {Lee}, {Li}, \& {Van Reeth}}]{saio2021}
{Saio}, H., {Takata}, M., {Lee}, U., {Li}, G., \& {Van Reeth}, T. 2021, \mnras, 502, 5856, \dodoi{10.1093/mnras/stab482}

\bibitem[{{Talon} \& {Charbonnel}(2003)}]{TC03}
{Talon}, S., \& {Charbonnel}, C. 2003, \aap, 405, 1025, \dodoi{10.1051/0004-6361:20030672}

\bibitem[{{Van Reeth} {et~al.}(2018){Van Reeth}, {Mombarg}, {Mathis}, {Tkachenko}, {Fuller}, {Bowman}, {Buysschaert}, {Johnston}, {Garc{\'\i}a Hern{\'a}ndez}, {Goldstein}, {Townsend}, \& {Aerts}}]{vanreeth18}
{Van Reeth}, T., {Mombarg}, J.~S.~G., {Mathis}, S., {et~al.} 2018, \aap, 618, A24, \dodoi{10.1051/0004-6361/201832718}

\bibitem[{{Vanon} {et~al.}(2023){Vanon}, {Edelmann}, {Ratnasingam}, {Varghese}, \& {Rogers}}]{riccardo3}
{Vanon}, R., {Edelmann}, P.~V.~F., {Ratnasingam}, R.~P., {Varghese}, A., \& {Rogers}, T.~M. 2023, \apj, 954, 171, \dodoi{10.3847/1538-4357/ace9db}

\bibitem[{{Varghese} {et~al.}(2023){Varghese}, {Ratnasingam}, {Vanon}, {Edelmann}, \& {Rogers}}]{ashlinmix23}
{Varghese}, A., {Ratnasingam}, R.~P., {Vanon}, R., {Edelmann}, P.~V.~F., \& {Rogers}, T.~M. 2023, \apj, 942, 53, \dodoi{10.3847/1538-4357/aca092}

\end{thebibliography}



\end{document}